\documentclass[aps,pra,10pt,notitlepage,twocolumn]{revtex4-2}
\usepackage{bbold,enumitem,mathtools}
\usepackage{tikz}
\usetikzlibrary{decorations.pathmorphing,patterns,decorations.markings,matrix,quantikz2}
\usepackage[export]{adjustbox}
\usepackage{nicematrix}
\usepackage{xstring}
\usepackage{ifthen}

    \usepackage{hyperref}
\usepackage[capitalise]{cleveref}

\usepackage[english]{babel}

\makeatletter
\def\bbl@set@language#1{%
  \edef\languagename{%
    \ifnum\escapechar=\expandafter`\string#1\@empty
    \else\string#1\@empty\fi}%
  \@ifundefined{babel@language@alias@\languagename}{}{%
    \edef\languagename{\@nameuse{babel@language@alias@\languagename}}%
  }%
  \select@language{\languagename}%
  \expandafter\ifx\csname date\languagename\endcsname\relax\else
    \if@filesw
      \protected@write\@auxout{}{\string\select@language{\languagename}}%
      \bbl@for\bbl@tempa\BabelContentsFiles{%
        \addtocontents{\bbl@tempa}{\xstring\select@language{\languagename}}}%
      \bbl@usehooks{write}{}%
    \fi
  \fi}
\newcommand{\DeclareLanguageAlias}[2]{%
  \global\@namedef{babel@language@alias@#1}{#2}%
}
\makeatother

\DeclareLanguageAlias{en}{english}

\newtheorem{example}{Example}
\newenvironment{proof*}[1][\proofname]{%
  
  \begin{proof}[#1]}{\end{proof}}
\newcommand{\half}{\mbox{$\textstyle \frac{1}{2}$}}

\newcommand{\identity}{\mathbb{1}}
\renewcommand{\epsilon}{\varepsilon}

\pgfset{
  foreach/parallel foreach/.style args={#1in#2via#3}{evaluate=#3 as #1 using {{#2}[#3-1]}},
}

\newcounter{arraycard}
\newcommand{\drawchain}[2][0.4\textwidth]{
\def\firstlist{#2,0}
\begin{center}
\begin{adjustbox}{width=#1}
\begin{tikzpicture}
\setcounter{arraycard}{0}
  \foreach \x in \firstlist {%
    \stepcounter{arraycard}
  }
  \foreach \x [count=\c]  in \firstlist
  {
    \ifthenelse{\c=\value{arraycard}}{}{\draw [thick] (2*\c,0) -- node[above,pos=0.5] {\x} (2*\c+2,0);}
    \node[circle,style={fill=black,minimum width=0.8cm,text=white}] at (2*\c,0) {};
  }

\end{tikzpicture}
\end{adjustbox}
\end{center}
}
\newcommand{\drawchainwithfields}[3][0.4\textwidth]{
\def\firstlist{#2,0}
\def\secondlist{#3}
\begin{center}
\begin{adjustbox}{width=#1}
\begin{tikzpicture}
\setcounter{arraycard}{0}
  \foreach \x in \firstlist {%
    \stepcounter{arraycard}
  }
  \foreach \x [count=\c,
  parallel foreach=\y in \secondlist via \c]
  in \firstlist
  {
  \IfInteger{\y}{
    \ifthenelse{\y=0}{}{
    \node[anchor=south] at (2*\c,0.4cm) {$\y$};
    }
    }{
  \node[anchor=south] at (2*\c,0.4cm) {\y};
    }
    \ifthenelse{\c=\value{arraycard}}{}{\draw [thick] (2*\c,0) -- node[above,pos=0.5] {\x} (2*\c+2,0);}
    \node[circle,style={fill=black,minimum width=0.8cm,text=white}] at (2*\c,0) {};
  }

\end{tikzpicture}
\end{adjustbox}
\end{center}
}

\begin{document}

\title{Incorporating Encoding into Quantum System Design}
\date{\today}
\author{Alastair \surname{Kay}}
\affiliation{Royal Holloway University of London, Egham, Surrey, TW20 0EX, UK}
\email{alastair.kay@rhul.ac.uk}
\begin{abstract}
When creating a quantum system whose natural dynamics provide useful computational operations, designers have two key tools at their disposal: the (constrained) choice of both the Hamiltonian and the the initial state of the system (an encoding). Typically, we fix the design, and utilise encodings \emph{post factum} to tolerate experimental imperfections. In this paper, we describe a vital insight that incorporates encoding into the design process, with radical consequences. This transforms the study of perfect state transfer from the unrealistic scenario of specifying the Hamiltonian of an entire system to the far more realistic situation of being given a Hamiltonian over which we had no choice in the design, and designing time control of just two parameters to still achieve perfect transfer.
\end{abstract}
\maketitle

\section{Introduction}

Quantum computers are at the stage of being able to perform some computations much faster than their classical counterparts, possibly even surpassing the requirements of quantum supremacy \cite{arute2019,madsen2022}. Nevertheless, these are very specific instances of algorithms, and we are still far from implementing arbitrary algorithms. That will need to wait until the available resources are significantly increased, and fault-tolerant computation becomes a reality. In the near-term, we instead operate with noisy, intermediate scale devices (the so-called `NISQ' era \cite{preskill2018}). A critical goal, then, is to implement all the elements of a computation as quickly and accurately as possible in order to maximise the quantum advantage of our device before being overcome by the inevitable decoherence. This means working at the fundamental, `machine code' level, which for many devices means describing the interactions between qubits with a Hamiltonian with time-controlled fields. For any given task, how quickly can it be implemented? How is that run-time affected by how much control we choose to implement? (Roughly speaking, the more control we use, the more potential to introduce error, so we want operations to be as fast as possible, but with as little control as possible.) How hard is it to find the controls that implement our desired operation?

One limit of this scenario is that of no control whatsoever, allowing the natural Hamiltonian dynamics to achieve the desired task. This requires a different, specific Hamiltonian for each task. One well-studied benchmark task in this context is known as perfect state transfer (PST) \cite{bose2003,christandl2004,christandl2005,kay2010a,karbach2005}, where one transfers an unknown quantum state between two remote regions of a quantum computer. Perhaps surprisingly, the results here give a faster transfer than that offered by the gate model of quantum computation \cite{yung2006}. The two-fold speed-up may be attributed to the use of multi-qubit interference instead of localised two-qubit operations. On the other hand, some minimal levels of control enable even faster transfer, saturating the limit imposed by the system's group velocity \cite{kay2009,osborne2004}.

Studies of PST have, since their inception, been charged with the major shortcoming of requiring precisely engineered conditions in order to achieve their results, being unable to adapt for manufacturing imperfections etc. High quality transfer \cite{apollaro2012,banchi2010} requires less variation in the coupling strengths, but still requires precision. This constraint has only been reduced in exchange for an uncertain arrival time \cite{burgarth2005a} or mitigated via the use of an encoding process \cite{haselgrove2005,kay2009,keele2021a,keele2021}.

\begin{figure}[!bt]
\centering
\includegraphics[width=0.35\textwidth]{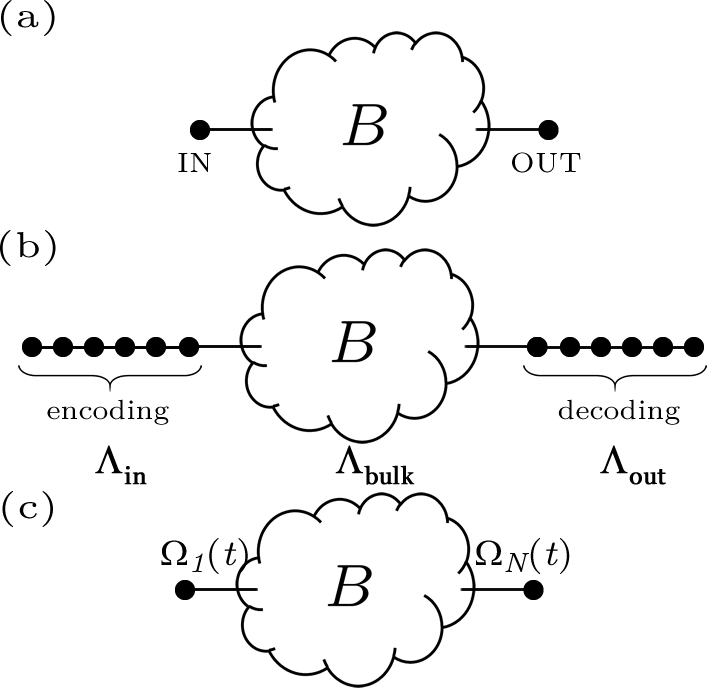}
\caption{A uniformly coupled spin chain, described by Hamiltonian $B$, is extended to fix some of the overall system's eigenvalues (b), permitting a task such as perfect encoded transfer between the two additions. (c) The extensions are simulated by time-varying control of two coupling strengths.}\label{fig:schematic}
\end{figure}

Our aim is to reject all such impositions, and instead demonstrate how an \emph{existing} system can be altered (extended) or controlled to achieve a task such as PST, as depicted in \cref{fig:schematic}. In this paper, we use a uniformly coupled system as proxy for no engineering requirements. A subsequent study \cite{kay2022a} will generalise the present results to near-universal applicability, the key difference being the need for a highly technical proof of the existence of solutions, which we are side-stepping by considering the uniform case. Thus, we are interested in a perfect quantum state transfer system where the central region is uniformly coupled, but we can choose the couplings at either end. One extreme, where one or two couplings at the end of the chain are chosen, is already known to give high quality, but imperfect transfer \cite{apollaro2012,banchi2010}. The opposite extreme is where all couplings may be chosen \cite{christandl2004} and results in perfect transfer. We bridge these two extremes, proving that when the central third of the chain is fixed, the error in the transfer is exponentially small in the chain length. This result is similar in nature to that of \cite{chen2016}, but with an exponential improvement in error behaviour. We will show that the same chain achieves perfect transfer if one uses the first and last thirds of the chain as encoding/decoding regions. Our method permits the creation of extensions of arbitrary lengths, and we numerically investigate the performance of these different extensions, and their trade off between transfer time and accuracy.

In \cref{sec:chains}, we review the required background of PST on spin chains, while also introducing our insight about how to use encoding methods. \cref{sec:extensions} shows how to symmetrically extend a pre-existing spin chain, fixing some of the eigenvalues of the overall system. These two methods combine to give perfect transfer. We explore the case of a chain, initially of 40 qubits, uniformly coupled, in \cref{sec:examples}. We extend this to a chain of 124 qubits on which perfect transfer can be observed. We consider the same chain from the perspective of end-to-end transfer, proving that high quality transfer results, approaching perfection exponentially quickly in the chain length. Finally, we numerically study the performance of shorter extensions to the original system, and find that they can also be extremely effective. In \cref{sec:create}, we extend the results beyond those of state transfer to the creation of useful, entangled, states.

We also apply a result of Haselgrove's \cite{haselgrove2005}, as depicted in \cref{fig:schematic}, which immediately demonstrates how to replace the extensions with time control of a single coupling strength at either end of the chain. Employing these results yields perfect communication through an imperfectly prepared system just by modifying the end couplings, and essentially maps to a constructive, analytic method of time control for perfect communication between two pendant vertices on a network, similar to the study of \cite{pemberton-ross2010}, studied from a control-theory perspective. The advantage of choosing the special case of a uniform chain to extend is that we have a good basis for comparison to results such as \cite{kay2009,murphy2010}.

Our methods rely heavily on those of Haselgrove \cite{haselgrove2005}. In that paper, two techniques were given: (i) for any given chain extension, how to find the optimal encoding (with no guidance about what the extension should be), and (ii) how to simulate an extension, replacing it with time control of a single coupling. The primary purpose of this paper is to extend the chain of reasoning, describing: (i) how to analytically identify when \emph{perfect} transfer is possible for a given extension and, as such, (ii) how to \emph{design} extensions that promise perfect transfer.

\section{Spin Chains}\label{sec:chains}

Consider a system Hamiltonian
\begin{equation}\label{eq:ham}
H=\half\sum_{n=1}^NB_nZ_n+\half\sum_{n=1}^{N-1}J_n(X_nX_{n+1}+Y_nY_{n+1}),
\end{equation}
in which $Z_n$ specifies a Pauli-$Z$ matrix applied to qubit $n$, and $\identity$ elsewhere. This describes a coupled chain of length $N$ with tunable coupling strengths $J_n$ and magnetic fields $B_n$. We limit ourselves to the field-free case of $B_n=0$ for simplicity. The $N$ qubits can be grouped into three distinct sets, $\Lambda_\text{in}$, $\Lambda_\text{bulk}$ and $\Lambda_\text{out}$, as depicted in \cref{fig:schematic}(b). Couplings between two qubits that are both in $\Lambda_\text{bulk}$ are assumed to be equal, and taken to be 1 without loss of generality. This is the Hamiltonian that we have been given and have no choice over. We retain the ability to choose the couplings on the extensions that we add to either end of the chain. We also take $\hbar=1$ so that all energies, transfer times etc.\ are dimensionless.

The Hamiltonian $H$ decomposes into subspaces characterised by the number of $\ket{1}$s in the basis elements. We focus on the single-excitation subspace, spanned by
$$
\ket{n}=\ket{0}^{\otimes (n-1)}\ket{1}\ket{0}^{\otimes(N-n)}.
$$
Within this subspace, we introduce the projectors onto the three different regions. For example,
$$
\Pi_\text{in}=\sum_{n\in\Lambda_\text{in}}\proj{n}.
$$

\emph{Perfect encoded state transfer} identifies a single-excitation state $\ket{\Psi_\text{in}}$ localised to the input region which evolves in the transfer time $t_0$ to $\ket{\Psi_\text{out}}=e^{-iHt_0}\ket{\Psi_\text{in}}$.
$$
\Pi_\text{in}\ket{\Psi_\text{in}}=\ket{\Psi_\text{in}},\quad\Pi_\text{out}\ket{\Psi_\text{out}}=\ket{\Psi_\text{out}}.
$$
This gives perfect transfer of a quantum state because an arbitrary superposition
$\alpha\ket{0}^{\otimes N}+\beta\ket{\Psi_\text{in}}$ can be created on the input region, evolving to the output state $\alpha\ket{0}^{\otimes N}+\beta\ket{\Psi_\text{out}}$.

Our primary goal is to discover how to choose the $\{J_n\}$ on the input and output regions such that we achieve perfect encoded state transfer. To assess the quality of transfer, we evaluate $\sigma$, the maximum singular value of $\Pi_\text{out}e^{-iHt_0}\Pi_\text{in}$, and define the fidelity to be $F=\sigma^2$ \cite{haselgrove2005} or transfer error $\epsilon=1-\sigma^2$. Note that, strictly, this is the excitation transfer fidelity. Given that the $\ket{0}$ component transfers perfectly, this is the worst-case fidelity for the transfer of an unknown quantum state. One could alternatively consider the average fidelity, $\frac13+\frac16(1+\sqrt{F})^2$, but we will always use this worst case $F$.

The left- and right-singular vectors in this case correspond to the states $\ket{\Psi_\text{out}}$ and $\ket{\Psi_{\text{in}}}$ respectively. End-to-end transfer is a special case with $\Pi_{\text{in}}=\proj{1}$ and $\Pi_{\text{out}}=\proj{N}$ \footnote{Viewed from this perspective, one does not need to artificially add a corrective phase, as is usually required in studies of state transfer.}.

Throughout this paper, we will assume symmetry: $J_n=J_{N-n}$. The reason for this is that the necessary and sufficient conditions for PST between opposite ends of a chain are well known \cite{kay2010a}, and include the requirement of this symmetry. While the symmetry is not necessary when one moves away from the end of the chain, it vastly reduces further requirements to a simple spectral condition. If we define the symmetry operator
$$
S=\sum_{i=1}^N\ket{N+1-i}\bra{i},
$$
then $SHS=H$ and hence, if transfer is perfect, $\ket{\Psi_\text{out}}= e^{i\phi}S\ket{\Psi_\text{in}}$ up to a known phase $\phi$ \footnote{This phase is known, and, for the field-free case that we consider in this paper, is just $\pi(N-1)/2$. It can be corrected by applying single-qubit phase gates on each qubit of the decoding region. We generally just consider this is incorporated into the decoding protocol where one has to map $\ket{\Psi_\text{out}}$ to the $\ket{1}$ of a single qubit.}. A symmetry operator $S_A$ is sized for a specific matrix $A$: $S_AAS_A=A$.

\subsection{Making use of Encoding}\label{sec:enc}

Our aim in this section is to reverse engineer the singular value description in order to guarantee perfect encoded state transfer. To that end, let the eigenvalues and eigenvectors of $H$ be $\lambda_n$ and $\ket{\lambda_n}$ respectively. These are ordered such that $\lambda_n>\lambda_{n+1}$. In a symmetric system, the Hamiltonian decomposes into two subspaces $H=H_+\oplus H_-$, with each eigenvalue being associated with one of these. Indeed, for a chain, $\lambda_{2n-1}\in\text{spec}(H_+)$ and $\lambda_{2n}\in\text{spec}(H_-)$ for all $n$. In such a symmetric system, the perfect transfer conditions using an encoded state $\ket{\Psi_\text{in}}$ at time $t_0$ are readily stated \cite{christandl2004}:
$$
\exists \phi: e^{-i\lambda_n t_0}=\pm e^{i\phi} \qquad\forall \lambda_n\in\text{spec}(H_\pm): \braket{\lambda_n}{\Psi_\text{in}}\neq0.
$$
Up to an arbitrary scale factor and offset, the spectrum for a PST system is a set of integers where the even (odd) integers are assigned to $H_+$ ($H_-$) and the perfect transfer time is $\pi$. In the field-free case of $B_n=0$, $\phi$ is an integer multiple of $\frac{\pi}{2}$.

For end-to-end transfer, where $\ket{\Psi_\text{in}}=\ket{1}$, then $\braket{\lambda_n}{\Psi_\text{in}}$ is non-zero for \emph{all} eigenvectors, and hence every eigenvalue must satisfy the integral condition. Transfer between nodes in the bulk of a chain has, to date, defied such a concise description because it is possible for a given $\braket{\lambda_n}{m}$ to be 0, meaning that the corresponding eigenvalue need not fulfil the spectral conditions. However, we will now use this to our advantage. Imagine that we have a fixed Hamiltonian $H$ whose eigenvalues we know. These may be categorised as the set $\Gamma_P$ which satisfy the perfect transfer conditions at time $t_0$, and $\Gamma_{\bar P}$, the imperfect ones which do not satisfy the perfect transfer conditions. If we can select an encoding $\ket{\Psi_\text{in}}$ such that for all $n\in\Gamma_{\bar P}$, $\braket{\lambda_n}{\Psi_\text{in}}=0$, we have perfect encoded state transfer. Our task is straightforward: find any state supported on $\Pi_\text{in}$ that is in the null space of
$$
\left\{\Pi_\text{in}\ket{\lambda}\right\}_{\lambda\in\Gamma_{\bar P}}.
$$
The existence of such a state is guaranteed provided the size of the encoding region is larger than $|\Gamma_{\bar P}|$.

\begin{example}\label{ex:first}
Consider the following chain:
\drawchain[0.45\textwidth]{$10\sqrt{5}$,$12\sqrt{14}$,$37\sqrt{6}$,$5\sqrt{185}$,$37\sqrt{6}$,$12\sqrt{14}$,$10\sqrt{5}$}
where each circle is a qubit, a number over an edge is a coupling strength $J$ between the specified pair of qubits, and a number over a qubit is a field strength $B_n$ on that qubit (and 0 if not specified). This system has eigenvalues $\sqrt{185}\times\{\pm1,\pm2,\pm6,\pm10\}$. We make the assignment
\begin{align*}
\Gamma_P&=\sqrt{185}\times\{\pm2,\pm6,\pm10\} \\
\Gamma_{\bar P}&=\{\pm\sqrt{185}\}.
\end{align*}
Using only the values in $\Gamma_P$, we have a perfect transfer time $t_0=\pi/(4\sqrt{185})$ since this gives the set of values
$$
\lambda_nt_0=\frac{-5\pi}{2},\frac{-3\pi}{2},\frac{-\pi}{2},\frac{\pi}{2},\frac{3\pi}{2},\frac{5\pi}{2},
$$
which have gaps of $\pi$ between them.

So, if we choose to take an encoding region of size $|\Gamma_{\bar P}|+1=3$, and evaluate the two eigenvectors of $\Gamma_{\bar P}$ restricted to the first 3 sites, we have
$$
\ket{1}\mp\frac{\sqrt{37}}{10}\ket{2}-\frac{3}{8}\sqrt{\frac{7}{10}}\ket{3}.
$$
There is a state
$$
3 \sqrt{7}\ket{1}+ 8\sqrt{10}\ket{3}
$$
that is orthogonal to both of these. In the time $t_0=\pi/(4\sqrt{185})$, this transfers to
$$
-i(3 \sqrt{7}\ket{8}+ 8\sqrt{10}\ket{6}),
$$
which is just on the decoding region (also size 3). We have perfect transfer of a single encoded excitation, and hence perfect transfer of a single encoded qubit.
\end{example}

By interpreting the use of the encoding/decoding regions in this way, we have the opportunity to incorporate encoding into our analytic strategies for the first time, rather than just adding it in subsequently. If the null space is of dimension $k$, one can encode $k$ qubits, and they will all be perfectly received (up to a suitable decoding sequence upon arrival) \cite{kay2010a}. Moving beyond the perfect transfer regime, a good use of the encoding strategy (although not necessarily optimal) is to find the eigenvalues that are the `worst offenders' (e.g.\ from the un-encoded case of just using the input $\ket{1}$) and set them to 0.

\section{Chain Extensions}\label{sec:extensions}

The application of this encoding strategy is now clear --- while PST requires engineering a system such that every eigenvalue satisfies a precise condition, we can forgo fixing $M$ eigenvalues in exchange for encoding/decoding regions of size $M+1$. We are now tasked with solving this problem: for a fixed central region, how do we symmetrically extend that chain such that it has certain eigenvalues of our choosing?

Our strategy is inspired by \cite{gesztesy1997}, which showed how to create a one-sided extension of a chain, fixing some of the eigenvalues. In the single excitation subspace, $H$ can be written in the form
$$
H=\begin{pNiceArray}{cwc{0.5cm}c|ccc|ccc}
\Block{3-3}{S_AAS_A}&&&&&&&&\\
&&&&&&&&\\
&&&J&&&&&\\\hline
&&J&\Block{3-3}{B}&&&&&\\
&&&&&&&&\\
&&&&&&J&&\\\hline
&&&&&J&\Block{3-3}{A}&&\\
&&&&&&&&\phantom{J}\\
&&&&&&&&
\end{pNiceArray}
$$
where $B=S_BBS_B$ is the fixed central region (i.e.\ a uniformly coupled chain), and $A$ is the output region that we will be able to select. The symmetry structure of $H$ becomes
$$
H_\pm=\left(\begin{array}{ccc|ccc}
& & & & & \\
&S_AAS_A& & & & \\
& & &J& & \\\hline
& &J& & & \\\
& & & &B_\pm & \\
& & & & &
\end{array}\right).
$$
Requiring a target eigenvalue $\lambda$ for $H$ with symmetry $\sigma\in\pm$ imposes that
$$
\text{det}(H_\sigma-\lambda\identity)=0.
$$
This can be expressed in terms of polynomials such as $Q_A(x)$, the characteristic polynomial of $A$, and $P_B^\sigma(x)$, the characteristic polynomial of the principal submatrix of $B_\sigma$, i.e.\ $B_\sigma$ with its first row and column removed.
\begin{equation}\label{eq:linear}
Q_A(\lambda)Q_B^\sigma(\lambda)=J^2P_A(\lambda)P_B^\sigma(\lambda).
\end{equation}
If $A$ comprises a set of $M$ sites, then $Q_A$ and $P_A$ are monic polynomials of degree $M$ and $M-1$ respectively. We don't know the coefficients of the polynomials, but there are $2M-1$ of them, and each instance of a $\lambda$ in \cref{eq:linear} is just a linear equation for those coefficients. Given $2M-1$ target eigenvalues, if a solution exists, $P_A(x)$ and $Q_A(x)$ are uniquely determined. Moreover, if we know $P_A(x)$ and $Q_A(x)$, we can uniquely reconstruct $A$ \cite{gladwell2005} (up to signs in the coupling strengths). In this argument, we have assumed that $J$ is known. Little change is required if $J$ is unknown, we just need one more parameter as, effectively, $P_A$ is no longer monic. 

\begin{example}
Starting from chain of 4 qubits,
\drawchain[0.26\textwidth]{$1$,$1$,$1$}
we demand a symmetric extension such that eigenvalues of the overall system include $\pm1$ and $\pm2$, the positive values being associated with the symmetric subspace.
\drawchain[0.48\textwidth]{$J_1$,$J$,$1$,$1$,$1$,$J$,$J_1$}
The symmetric subspace is also equivalent to a chain, but with a non-zero field on the final spin:
\drawchainwithfields[0.26\textwidth]{$J_1$,$J$,$1$}{0,0,0,1}
We impose that the symmetric subspace should contain the eigenvalues $1,2$. Dividing this into two sections
$$
A=\left(\begin{array}{cc} 0 & J_1 \\ J_1 & 0 \end{array}\right),\qquad
B_+=\left(\begin{array}{cc} 0 & 1 \\ 1 & 1 \end{array}\right)
$$
then we can readily evaluate
$$
\frac{Q_B^+(x)}{P_B^+(x)}=\frac{x(x-1)-1}{x-1}
$$
which must equal
$$
\frac{J^2P_A(x)}{Q_A(x)}=\frac{J^2x}{x^2-J_1^2}
$$
at $x=1,2$.
In this case, we directly solve the two simultaneous equations to find
$J_1^2=1$ and $J^2=\frac32$.
We thus see that the chain
\drawchain[0.45\textwidth]{$1$,$\sqrt{\frac32}$,$1$,$1$,$1$,$\sqrt{\frac32}$,$1$}
has the desired eigenvalues.
\end{example}

In this section, we have shown how to take a fixed central region contains $M_B$ qubits, extending it to have $N=2M+M_B$ qubits, with the ability to fix $2M$ eigenvalues (in the $J$ unknown case). Once we incorporate our conclusions of \cref{sec:enc} about the use of encoding, perfect encoded transfer is possible provided $M>M_B$.

\section{Examples}\label{sec:examples}

For the purposes of numerical examples, it is convenient to use the field-free case, i.e.\ where the matrices $A$ and $B$ have 0 on the diagonal, halving the number of parameters we have to work with. We shall assume that $B$ comprises an even number of qubits, such that the chain as a whole has an even number of qubits. As a result, both $B$ and $H$ will have all eigenvalues occurring in $\pm\lambda$ pairs. Note that $\ket{\lambda}$ and $\ket{-\lambda}$ have opposite symmetries. Instead of solving \cref{eq:linear} for both, we can build this feature into the structure of the polynomials that we're solving for, specifically that $Q_A(x)$ comprises only even (odd) powers if $M$ is even (odd), while $P_A(x)$ is the opposite.

Solving the linear equations (\ref{eq:linear}) directly is challenging as the structures involved closely resemble Vandermonde matrices, including terms such as $\lambda^M$, which rapidly lead to numerical instabilities. Instead, we recognise that the problem is that of finding a rational function of specific degrees which fits known values at specific points. There are several existing techniques for solving this such as Thiele's continued fraction routine \cite{stoer2013}. In all our numerical tests, we have used Algorithm 1 presented in \cite{egecioglu1989} (see also \cite{gemignani1993}) as a particularly efficient algorithm whose iterative structure will be familiar to those who work with tridiagonal matrices or orthogonal polynomials. However, this only works when $J$ is unknown \footnote{The algorithm is not able to trade the loss of one parameter (a target eigenvalue) for the gain of another (the known $J$), as it chooses to reduce the degree of $P$ by 1 instead of causing it to be monic. However, when, as here, we restrict to a field-free scenario, there is no difference between a case of fixed $J$ and initial chain length $M$, and unknown $J$ and chain length $M+2$.}.

To incorporate the field-free assumption into the rational interpolation algorithm \cite{egecioglu1989}, we need the function $f(x)=\frac{P(x)}{Q(x)}$ to be anti-symmetric in the case of $A$ being of even length. If we have positive points (target eigenvalues) $x_i$ for which the rational function must have values $f_i$ (and hence also values $-x_i$ such that $f(-x_i)=-f_i$), then instead we attempt to find a rational function $g(x)$ which satisfies $\{g(x_i^2)=f_i/x_i\}$. This means that we will have determined
$$
g(x)=\frac{p(x)}{q(x)},
$$
from which we can construct
\begin{equation}
f(x)=\frac{xp(x^2)}{q(x^2)}=\frac{P(x)}{Q(x)}.\label{eq:sub}
\end{equation}
It is straightforward to verify the required relations
\begin{align*}
f(x_i)&=x_ig(x_i^2)=f_i \\
f(-x_i)&=-x_ig(x_i^2)=-f_i.
\end{align*}

\begin{figure}
\centering
\includegraphics[width=0.45\textwidth]{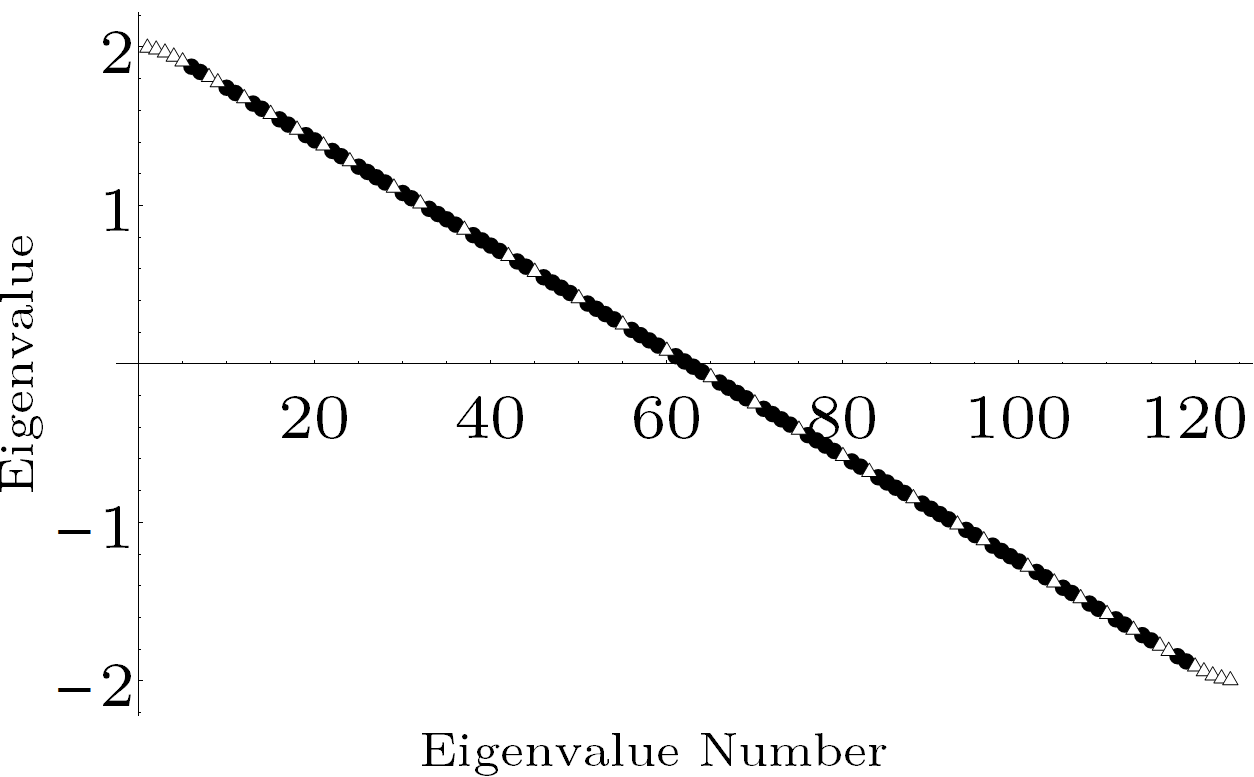}
\caption{Extending a uniformly coupled chain of 40 qubits to 124 qubits, achieving perfect encoded transfer. We chose certain eigenvalues (circles) to fit the perfect transfer conditions. Even the unfixed ones (triangles) gave a good approximation to those conditions.}\label{fig:uniform_distrib}
\end{figure}

\subsection{Perfect Encoded Transfer}\label{sec:encodedtransfer}

\begin{figure*}
\centering
\includegraphics[valign=t,width=0.45\textwidth]{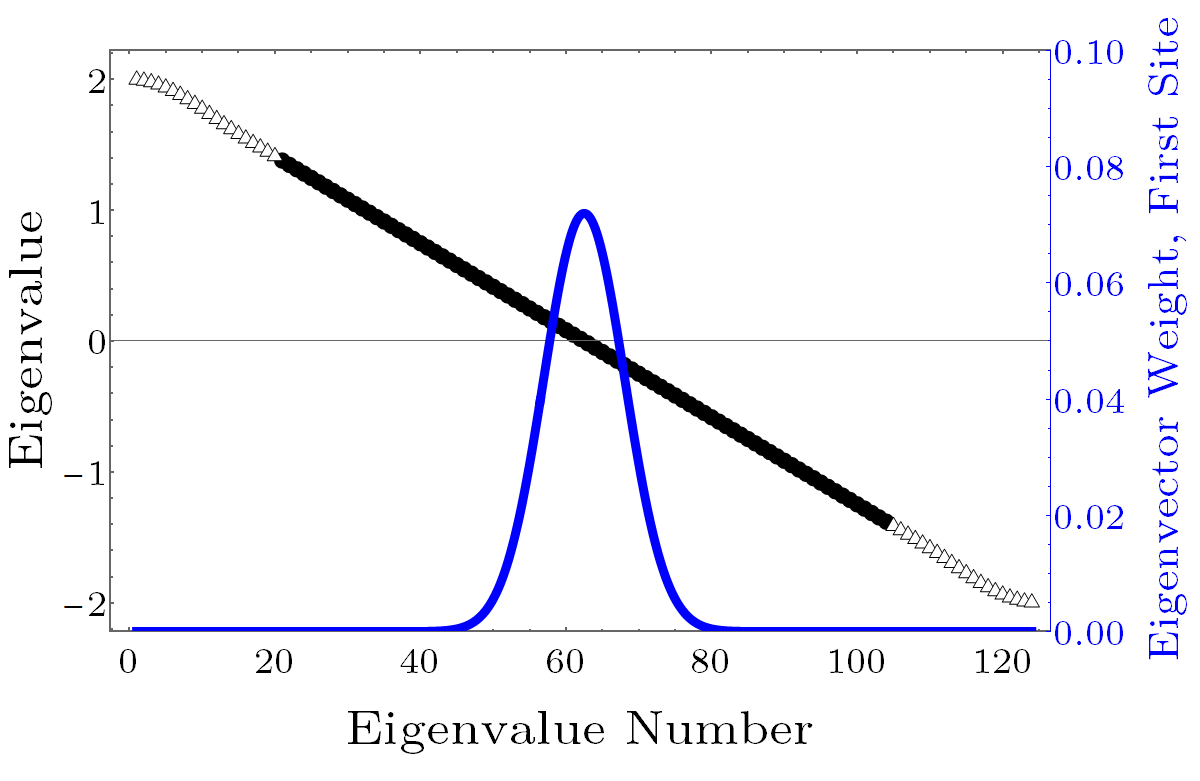}
\hspace{0.5cm}
\includegraphics[valign=t,width=0.45\textwidth]{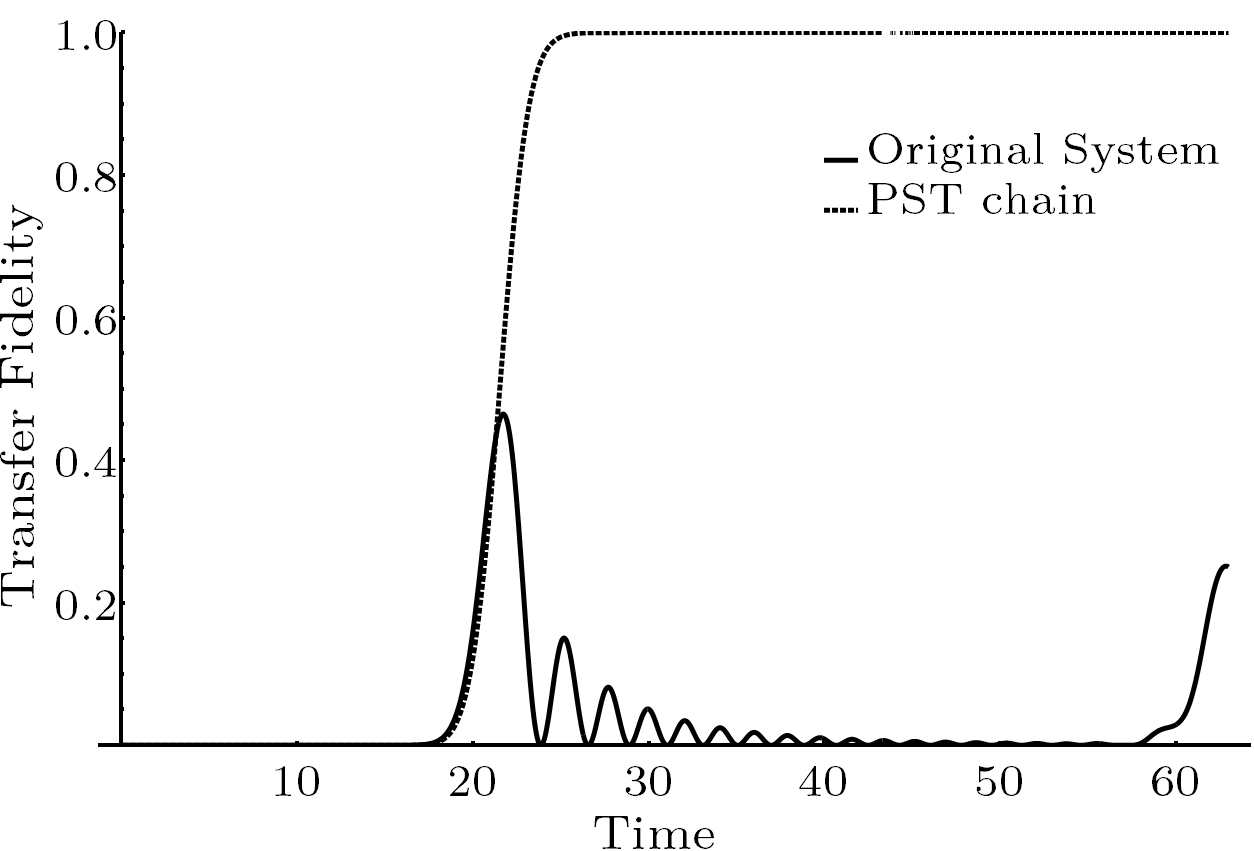}
\caption{(a) Eigenvalues of a chain extended from 40 qubits to give PST. The central eigenvalues satisfy the linear condition (circles), with the others (triangles) being uncontrolled. The continuous (blue) line depicts the weight of each eigenvector on the first site of the chain. (b) Plot of the encoded state transfer fidelity of the extended perfect transfer chain (dashed) compared to original chain (continuous). A plot of the encoded state transfer fidelity using a 124 qubit uniformly coupled chain is indistinguishable from the perfect transfer case at the resolution of this image (\cref{fig:smaller_extensions} shows how this disparity changes with extension length).}\label{fig:uniform_distrib_central}
\end{figure*}

Consider the example where the central region comprises 40 qubits, uniformly coupled, strength 1. We will introduce $M$ qubits (even) at either end of the chain, and fix a value $\delta$ to be as large as possible, corresponding to a state transfer time of $t_0=\pi/\delta$. We seek couplings of the extension such that the overall system has eigenvalues
\begin{equation}
\delta\left\{M-2k-\half\right\}_{k=0}^{M-1} \label{eqn:spectrum}
\end{equation}
in the symmetric subspace. A field-free system must also have eigenvalues $\delta\left\{M-2k-\frac32\right\}$ in the antisymmetric subspace. Setting $M=42$ is sufficient to guarantee perfect transfer, assuming a solution exists. While we have no guarantee about the existence of a solution (which is beyond the scope of this paper, see \cite{kay2022a}), this particular model is quite forgiving. Recall that the spectrum for a uniformly coupled chain of $N$ qubits is
$$
2\cos\left(\frac{\pi n}{N+1}\right).
$$
In the central region $n\sim\frac{N+1}{2}$, the spectrum is near-linear. Thus, imposing that it should be exactly linear in its central region is not a big deviation.What value should be chosen for the gradient, $\delta$? Two extremes yield a tight range to numerically search within for the optimal:
\begin{itemize}
\item A uniformly coupled system has gradient $\frac{2\pi}{N+1}$ in the central region of its spectrum, and this has the fastest possible group velocity (i.e.\ transfer speed) of any system with maximum coupling strength $J_{\max}=1$. Given that we are fixing about 2/3 of the eigenvalues, which is much more than the typical linear region, it seems unlikely that we will be able to match this gradient.
\item The fastest perfect transfer system \cite{christandl2004} has gradient $\frac{4}{N}$ throughout its spectrum (and is approximately uniform in the central region). A solution matching this gradient should exist. This extreme may suggest the possibility of even getting the eigenvalues that we don't choose close to the linear pattern as well. See \cref{fig:uniform_distrib} for an example.
\end{itemize}
\Cref{fig:uniform_distrib_central}(a) depicts our chosen example where we have focussed on fixing the eigenvalues in the central region, as specified by \cref{eqn:spectrum}. The code used to produce this example can be found at \cite{kay2023}, including the explicit solutions for the coupling strengths. This achieves perfect encoded transfer in time 94.5 (as compared to the PST gradient giving transfer time $\sim 99$). However, by using the optimal encoding \cite{haselgrove2005}, we see in \cref{fig:uniform_distrib_central}(b) that extremely high fidelity transfer is achieved in a much shorter time, essentially coinciding with the first arrival peak of a wavepacket travelling at the maximum group velocity of the system, i.e.\ as fast as transfer could possibly occur. We comment on why this is the case in \cref{sec:encodedtime}.

\begin{figure}
\centering
\includegraphics[width=0.45\textwidth]{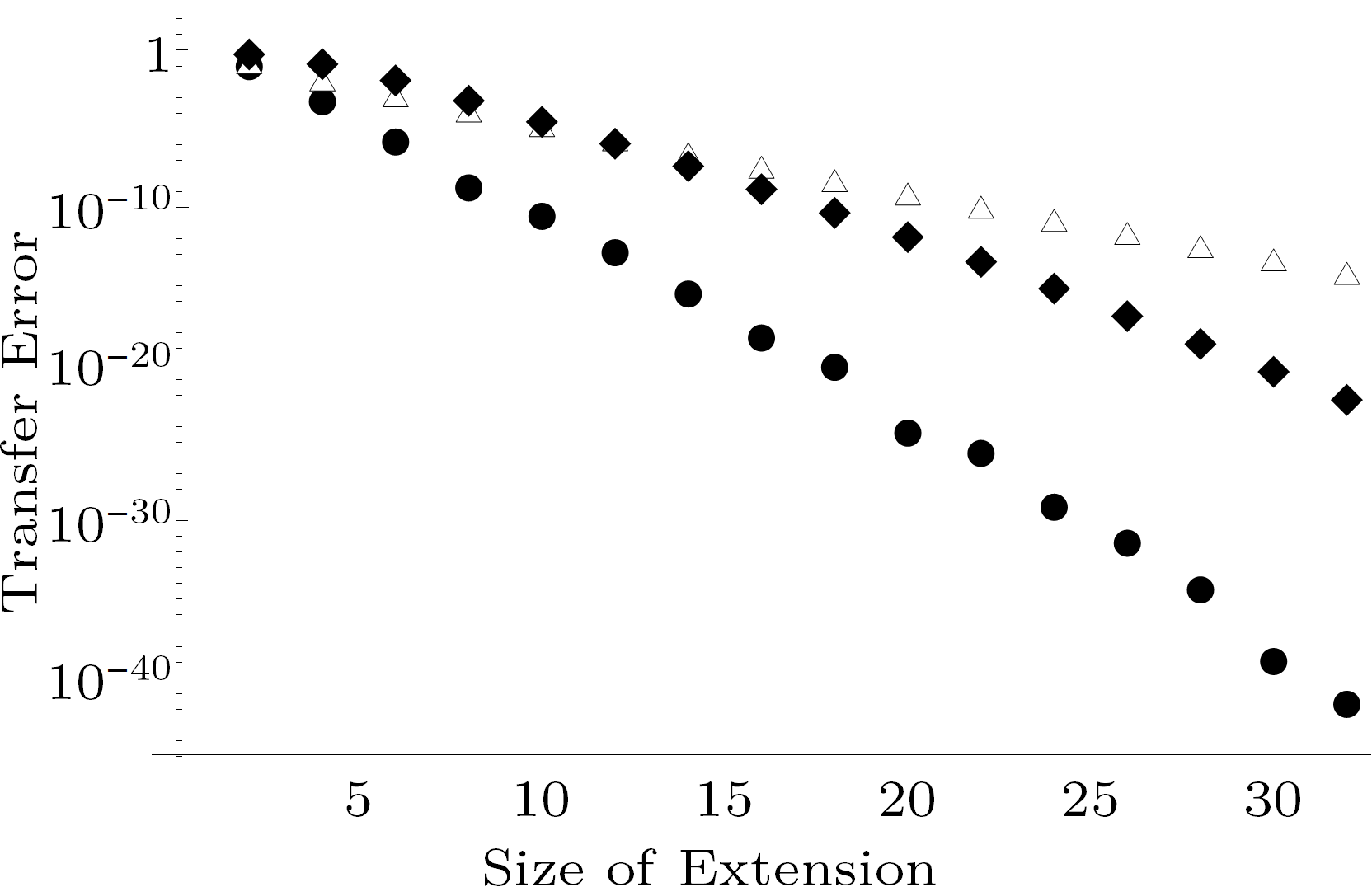}
\caption{For a central region of 40 qubits, we find symmetric extensions and assess the transfer error at $t=\pi/\delta$ where $\delta$ is fixed. We compare end-to-end transfer (triangles) and optimal encoding over the entire encoding/decoding region (circles). We also include the time-optimised transfer error of a uniformly coupled chain of the same length (diamonds) using encoding over the entire encoding/decoding region.}\label{fig:smaller_extensions}
\end{figure}

We should compare this solution to the best known previous solution, in which one simply extends the uniform chain with another uniform chain, and uses the optimal encoding, which is inspired by creating wavepackets that travel through the system at the group velocity \cite{osborne2004,kay2009}. Visual inspection reveals that the solutions are comparable. For the uniform chain, however, transfer is never perfect, and as soon as the central region is not uniformly coupled, we don't know how to proceed. Nevertheless, this comparison suggests that it might be interesting to reduce the size of the controlled regions. In \cref{fig:smaller_extensions}, we see that even modestly sized extensions, supplemented by encoding, are extremely effective in improving transfer fidelity \footnote{For smaller extensions, it should be possible to increase $\delta$. However, we kept it fixed so that all systems had the same perfect transfer time.}, indeed far more effective than uniform extensions.

\subsection{Imperfect State Transfer Families without Encoding}\label{sec:noenc}

\Cref{fig:uniform_distrib_central} shows one further striking feature. It plots the weight of each eigenvector on the first site, and this is strongly weighted in the central region where the eigenvalues have been tuned to have the linear relation required of perfect transfer. Thus high quality transfer will result \emph{without any encoding}.

To study this in greater detail, let us assume that a family of solutions of the form described in \cref{fig:uniform_distrib_central,eqn:spectrum} exist. Let $\ket{\psi}=\sum_na_n\ket{\lambda_n}$ be our encoding, giving a decoding of $\ket{\phi}=\sum_na_n(-1)^{n+1}\ket{\lambda_n}$. Now let $\Gamma_{\bar P}$ be the set of indices $n$ for which the $\lambda_n$ do not satisfy the PST conditions. At worst case, the transfer fidelity would be
$$
F_{\min}=1-2\sum_{n\in\Gamma_{\bar P}}|a_n|^2.
$$
In Fig.\ \ref{fig:uniform_distrib_central}, the spectrum is very close to linear throughout its range (and exactly linear in the central region). The energy gap must be approximately $\delta=4J/(N-1)$, giving a transfer time of $\pi/\delta$. Since the spectrum entirely determines the values $a_n$ on a symmetric chain, this conveys that the $a_n$ will be extremely close to those of the perfect transfer chain \cite{christandl2004,albanese2004}, so we can take the analytic solutions for those eigenvectors as excellent approximations. Thus, for a chain of total length $N$,
$$
|a_n|^2=\frac{1}{2^{N-1}}\binom{N-1}{n-1}.
$$
We take the large $N$ limit, so the summation for $F_{\min}$ becomes an integral. With $\Gamma_{\bar P}$ being all those eigenvalues apart from the central $2N/3$ of them, we get an error of
$$
1-F_{\min}^{\text{PST}}=\frac{8}{\sqrt{2\pi}}\int_{-\infty}^{-\sqrt{N}/6}e^{-2\theta^2}d\theta<\frac{12}{\sqrt{2N\pi}}e^{-N/18}.
$$
This yields asymptotically perfect transfer between opposite ends of the chain. This is an exponential improvement in approach compared to \cite{chen2016}. It might be considered to be taking the studies of \cite{apollaro2012,banchi2010} to their ultimate limit, demonstrating how many couplings it is sufficient to fix in order to get asymptotically perfect transfer with a uniformly coupled central region and end-to-end transfer, not just a fidelity over some finite threshold.

From our proof of end-to-end transfer, it is also clear that for any $(N-|\Gamma_{\bar P}|)/2$ that grows faster than $\sim\sqrt{N}$, the integral will also vanish. So perhaps we only need to fix $O(\sqrt{N})$ eigenvalues? This is a different proposition as, with control over fewer eigenvalues, it is less likely that we approximate the linear spectrum for its full range. The spectrum will be much closer to that of the uniform chain, more of whose eigenvectors have non-negligible support on the first site. One might expect to compensate with encoding. Fig.\ \ref{fig:smaller_extensions} shows how the transfer fidelity varies with the length of the extension, with the error dropping exponentially. For example if we have only extended by 8 spins, rather than the $\sim 40$ required to achieve perfect transfer, we still achieve a transfer with error approximately $2\times 10^{-9}$, while the uniformly extended chain only achieves an error of $10^{-4}$.


\subsection{Time for Encoded Transfer}\label{sec:encodedtime}

The fact that these chains permit extremely high quality end-to-end transfer, being closely related to a PST chain, yields some insight about the speed of the high accuracy encoded state transfer, observed in \cref{fig:uniform_distrib_central}(b). Consider a PST chain of length $N$ such as in \cite{christandl2004}, with a state transfer time $t_0$ (which grows linearly in $N$ if we scale the system such that its maximum coupling strength is 1). Starting localised at the first site, the motion is essentially the ballistic motion of a wavepacket, centred on the position
$$
(N-1)\sin^2\left(\frac{\pi t}{2t_0}\right)+1
$$
with a spread
$$
\sigma=\frac{\sqrt{N-1}}{2}\sin\left(\frac{\pi t}{t_0}\right).
$$
In fact, the distribution of probabilities per site is exactly that of an $(N-1)$-sample Bernoulli distribution with $p(t)=\sin^2\left(\frac{\pi t}{2t_0}\right)$.
Thus, the wavepacket is almost entirely restricted to the encoding region until a time $t_{\text{in}}$ (such that $p(t_{\text{in}})\sim \frac13$) and almost entirely restricted to the decoding region after a time $t_0-t_{\text{in}}$. When the wavepacket is in those regions, we can recreate it with encoding and decoding. The resulting error for encoding may be bounded by a Chernoff bound as
$$
\epsilon\leq e^{-(N-1)p(3p-1)^2/(21-9p)}
$$
with a symmetric equivalent for the decoding error. Hence, we only need a transfer time of approximately $0.22t_0+O(1/\sqrt{N})$ with an error that is exponentially small in $N$. This time, $0.17N$, compares extremely well with the limit of $N/6$ imposed by the maximum group velocity being 2 \footnote{Note also that this calculation was using the standard PST time from \cite{christandl2004}. We observed in \cref{sec:encodedtransfer} that solutions exist for a slightly higher gradient, and hence shorter transfer time.}. While this strategy may not be optimal, it provides a lower bound for the performance. We should note however that the spread of the wavepacket, $O(N^{1/2})$, is broader than the optimal wavepacket for the equivalent uniformly extended chain, $O(N^{1/3})$ \cite{osborne2004}.

\section{Simulating Extensions with Time Control}\label{sec:state_creation}

The work of \cite{haselgrove2005} contains two useful strategies. We have made extensive use of one here --- the encoding/decoding of the state, giving it a new interpretation for how it can be used to achieve perfect transfer. The second, how certain sections of a spin system can be replaced by time control, is just as useful. It can be used directly, without alteration. Instead of adding many qubits to the initial fixed system, we just control (varying in time) two of the coupling strengths. This is depicted conceptually in Fig.\ \ref{fig:schematic}. Transfer at fidelity $F$ between encoding and decoding regions translates directly to transfer between the two extremal sites in the virtualised system at fidelity $F$.

Specifically, if we have a system for which there's a chain that we want to simulate, let the qubits of the chain be indexed $1$ to $M$, where $M$ is the end of the chain, and 1 is connected to the rest of the system. The evolution of the single excitation that we want to simulate has amplitudes $\psi_n(t)$ at site $n$. We can assume that $\psi_2(t)$ is real. We can remove all the qubits 3 to $M$, and replace the coupling $J_{1,2}$ with
$$
\Omega(t)=\frac{J_{1,2}\psi_2(t)}{\sqrt{\sum_{n=2}^M|\psi_n(t)|^2}}.
$$
\begin{example}
Consider the following chain from \cref{ex:first}:
\drawchain[0.45\textwidth]{$10\sqrt{5}$,$12\sqrt{14}$,$37\sqrt{6}$,$5\sqrt{185}$,$37\sqrt{6}$,$12\sqrt{14}$,$10\sqrt{5}$}
We have already determined an encoding
$$
3 \sqrt{7}\ket{1}+ 8\sqrt{10}\ket{3}
$$
and decoding
$$
-i(3 \sqrt{7}\ket{8}+ 8\sqrt{10}\ket{6})
$$
after time $t_0=\pi/(4\sqrt{185})$ that achieve perfect encoded transfer. We can calculate how each of the amplitudes evolves in time. For example,
$$
\psi_1(t)=3 \sqrt{\frac{7}{703}} \cos^5\left(2 \sqrt{185} t\right).
$$
Hence, we can replace the extensions with time controls
\drawchain[0.23\textwidth]{$\Omega_1(t)$,$5\sqrt{185}$,$\Omega_N(t)$}
with
\begin{align*}
\Omega_1(t) &=\frac{2\sqrt{555} \left(29-45 \cos \left(4 \sqrt{185}
   t\right)\right)}{\sqrt{535-702 \cos \left(4 \sqrt{185}
   t\right)+243 \cos \left(8 \sqrt{185} t\right)}}\\
\Omega_N(t) &=\frac{2\sqrt{555} \left(29+45 \cos \left(4 \sqrt{185}
   t\right)\right)}{\sqrt{535+702 \cos \left(4 \sqrt{185}
   t\right)+243 \cos \left(8 \sqrt{185} t\right)}}.
\end{align*}
Using this, one excitation transfers from the first qubit to the last in time $t_0$. Moreover, any initial state $\alpha\ket{0}+\beta\ket{1}$ transfers between those two spins in the same time.
\end{example}
There are several features of the above example which are worth noting: (i) the symmetry in the time control, $\Omega_N(t)=\Omega_1(t_0-t)$; (ii) the boundedness of the control fields, $|\Omega|\leq J$; (iii) the smoothness of the control fields.

\subsection{Creation of States}\label{sec:create}

In fact, the ability of the virtualisation technique of \cite{haselgrove2005} to simulate a perfect transfer chain in which every site perfectly transfers to its mirror site in the perfect transfer time $t_0$ has some extremely powerful consequences. We will now use this to produce arbitrary states of a single excitation on a chain.

Imagine that we're given a uniformly coupled chain of $M_B$ qubits. Before appending engineered chains for the purpose of tuning the spectrum, we add a further $M_B$ uniformly coupled qubits. We will refer to these as the ``mirror system''. Then we add the extra chains at both ends to tune the spectrum. The longer the chains, the more accurate the protocol that we'll realise, at the cost of longer time. Once we've solved for that system, we will virtualise everything that we've added.

Note that the virtualisation procedure is state dependent, being derived from the amplitudes $\{\psi_n(t)\}$ of the single excitation's evolution in the original system: it depends on the system's initial state. It is this dependence on the initial state that we will now utilise. In particular, if we create \emph{any} single excitation state that we like in the mirror system, then in the PST time, it arrives perfectly on the original system, while the virtualisation procedure will just reduce the initial state to a single excitation on the qubit that replaces the chain section. A numerical example is explicitly given in \cite{kay2023}. Translated into the virtualised system, you start with a single excitation on one of the two extremal spins, and the time control determines any single-excitation output state that you desire on the bulk system!

We can illustrate these concepts using a PST chain \cite{christandl2004}.
\begin{example}
Consider a PST chain of 6 qubits,
\drawchain[0.4\textwidth]{$\sqrt{5}$,$\sqrt{8}$,$3$,$\sqrt{8}$,$\sqrt{5}$}
Pick any initial state localised on qubits 4 to 6 (the mirror of qubits 1 to 3), such as $(\ket{4}-\sqrt{10}\ket{6})/\sqrt{11}$. In time $t=\pi/2$, this state is perfectly mirrored onto qubits 1 to 3: $i(\sqrt{10}\ket{1}-\ket{3})/\sqrt{11}$. We can virtualise this using the shorter chain
\drawchain[0.26\textwidth]{$\sqrt{5}$,$\sqrt{8}$,$\Omega(t)$}
with a time-varying coupling strength
$$
\Omega(t)=\frac{3\sqrt{2} (2-\cos (2 t))}{\sqrt{2 \cos (2 t)-\cos (4 t)+21}}.
$$
The virtualised system now undergoes the evolution
$$
\ket{4}\xrightarrow{\ t=\pi/2\ }i(\sqrt{10}\ket{1}-\ket{3})/\sqrt{11}.
$$
Any input of the form $\alpha\ket{0}+\beta\ket{4}$ evolves, under the same time control, to
$$
\xrightarrow{\ t=\pi/2\ }\alpha\ket{0000}+i\frac{\beta}{\sqrt{11}}(\sqrt{10}\ket{1}-\ket{3}).
$$

Different time controls, such as $\Omega(t)=$
$$
\frac{3\sqrt{10} (1-2 \cos (2 t)+3 \cos (4 t))}{\sqrt{105-98 \cos
   (2 t)+76 \cos (4 t)-54 \cos (6 t)+27 \cos (8 t)}}
$$
yield different evolutions,
$$
\ket{4}\xrightarrow{\ t=\pi/2\ }-i(\sqrt{2}\ket{1}+\sqrt{5}\ket{3})/\sqrt{7}
$$
for the same initial state.
\end{example}

\begin{figure}
\centering
\includegraphics[width=0.45\textwidth]{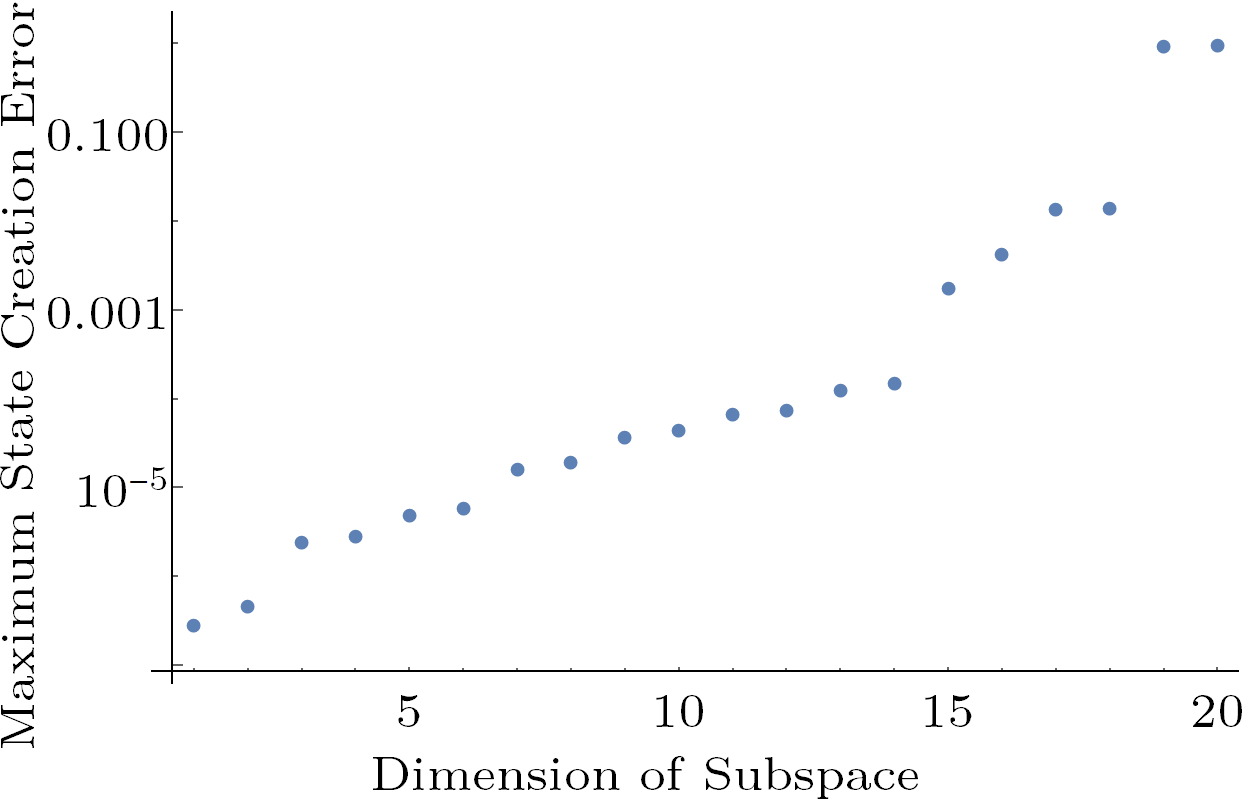}
\caption{For a 40 qubit chain as specified in \cref{fig:uniform_distrib}, states can be created on the first 20 qubits of the bulk with a maximum error for a given dimension of space.}
\label{fig:state_creation}
\end{figure}

Of course, our strategy will only ever approximate a perfect transfer chain. To that end, we take the state $\ket{\Psi}$ that we want to create, and run the Hamiltonian evolution backwards to find the best possible starting state. If $\Pi_{\text{out}}$ is the projector onto the output region, including also the mirror system, then up to normalisation, the best possible starting state is
$$
\ket{\Psi_{\text{in}}}=\Pi_{\text{out}}e^{iHt_0}\ket{\Psi},
$$
and the transfer fidelity is
$$
\bra{\Psi}e^{-iHt_0}\Pi_{\text{out}}e^{iHt_0}\ket{\Psi}.
$$
In order to understand the efficacy of our system, let us calculate the eigenvalues
$$
\Pi_{\text{bulk}}e^{-iHt_0}\Pi_{\text{out}}e^{iHt_0}\Pi_{\text{bulk}}.
$$
For the chain specified by \cref{fig:uniform_distrib}, these are plotted in \cref{fig:state_creation}. We see that there's a large space from which states can be created with high fidelity.

\section{Conclusions \& Future Work}

We have shown how a uniformly coupled chain can be symmetrically extended by $M$ qubits on either side, fixing $2M-1$ (or $2M$) of the eigenvalues to those that we specify. By also implementing an encoding/decoding procedure over the $M$ qubits at either end, we can avoid populating up to $M-1$ eigenvectors whose eigenvalues do not satisfy the perfect transfer condition. We can thus create a perfect encoded transfer chain where the central third is fixed to being uniformly coupled. Operating close to the speed limit of the system yields a transfer whose error is exponentially small in the chain length. Moreover, thanks to \cite{haselgrove2005}, all the additional spins can be `virtualised', i.e.\ replaced simply by time control of a single coupling strength at either end of the uniform chain. We have demonstrated numerically that even with shorter extensions, extremely high fidelity transfer can be achieved. Equally, if one wants to dispense with encoding, high quality transfer is possible, with an error that decreases exponentially in the chain length. A small modification of the protocol allows for the creation of a wide range of single-excitation states. The algorithms for computing the extensions, and corresponding time control in the virtualisation, are extremely efficient.

The formalism developed here is not limited to the initial system being a uniformly coupled chain. \emph{Any} coupling topology and specification of coupling strengths is equally amenable. However, the challenge is ensuring that solutions to the set of linear equations (\ref{eq:linear}) exist. That, and the consequences for transfer speed, are topics for a future paper \cite{kay2022a}. IBMQ devices of various geometries are thus a promising avenue for experimental realisation: they already use a Hamiltonian of the form \cref{eq:ham}. Moreover, since we are already using encodings, these encodings can be optimally updated to incorporate knowledge of the system noise \cite{keele2021a}. As such, this methodology heralds a new era for quantum state transfer, and related studies, in which we can adapt to a provided system rather than having to request specific properties.

That said, there remain limitations. The most obvious ones are that (i) systems such as IBMQ do not directly provide access to time control of coupling strengths, only the local magnetic fields, (ii) if the extended system is a chain, multiple excitations behave well \cite{kay2010a}. However, if the virtualisation procedure of \cite{haselgrove2005} is used, the single excitation subspace no longer provides a good description of the behaviour in higher excitation subspaces, and (iii) we don't yet know how to incorporate the treatment of noise such as \cite{keele2021a} into the virtualisation procedure. These are issues that we hope may be addressed in the future.


%

\end{document}